\documentclass[12pt]{article}
\usepackage[latin1]{inputenc}
\usepackage{amssymb}
\usepackage{subfigure}
\usepackage{graphicx}
\usepackage{float}

\usepackage{amsmath}
\usepackage{color}

\newcommand{\bb}{\begin{equation}}
\newcommand{\ee}{\end{equation}}
\newcommand{\bega}{\begin{eqnarray}}
\newcommand{\ega}{\end{eqnarray}}
\newcommand{\begae}{\begin{eqnarray*}}
\newcommand{\egae}{\end{eqnarray*}}

\newcommand{\h}{\hspace*{4ex}}

\newcommand{\cent}{\centerline}
\newcommand{\vs}{\vspace*}

\begin{document}

\baselineskip 0.5cm

\begin{center}

{\large {\bf Generation and characterization of an array of Airy-vortex beams } }%$^{\: (\dag)}$ } 

\end{center}

\vs{0.2 cm}

\cent{Rafael A. B. Suarez$^{\: 1}$, Antonio A. R. Neves$^{\: 1}$, and Marcos R. R. Gesualdi$^{\: 1}$}

\vs{0.2 cm}

\centerline{{\em $^{\: 1}$ Universidade Federal do ABC, Av. dos Estados 5001, CEP 09210-580, Santo Andr\'e, SP, Brazil.}}

\vs{0.5 cm}

{\bf Abstract  \ --} \ We present the experimental generation and analysis of both the intensity and phase of an optical vortex beam originating from a superposition of Airy-vortex beams. A new theoretical proposal is accessible to generate an array of Airy-vortex beams (AiVBs) from a superposition of N identical AiBs symmetrically located with respect to the propagation axis, where each beam is superimposed with OV and are characterized by the same topological charge l. Additionally, a single experimental apparatus enables both the optical generation and reconstruction of computer-generated holograms implemented on a spatial light modulator and its phase analysis by digital holography. The experimental results agreed with the theoretical predictions. The array of Airy-vortex beams has potential applications for trapping and rotating micro-particles in optical tweezers, modulation in optics communications, optical metrology, and others in optics and photonics. \\

%\h PACS nos.: 42.25.Bs; 42.25.Fx; 41.20.Jb; 46.40.Cd; 41.85.-p; 46.40.Cd; 42.20.Ht; 42.40.Jv, 42.30Lr.

\vs{0.5 cm}

\h {\em\bf 1. Introduction} --- An optical vortex (OV) is characterized by a helical phase structure that carries orbital angular momentum, whose magnitude depends on the topological charge $l$. This phase is superimposed onto the beams which lead to a phase singularity at the center of the radiation field, where the phase of the beams is undetermined and its amplitude vanishes \cite{Allen1992}. These features have broad application in different areas, including optical trapping \cite{Paterson2001,Dholakia2011}, optics communication \cite{Willner2015}, optical metrology and quantum information \cite{Molina2007}. So far, several optical vortexes have been superimposed and reported with other types of beams, such as Laguerre Gaussian vortex beams \cite{Flossmann2005}, Bessel-vortex beams \cite{Orlov2002}, and Airy-vortex beams \cite{Mazilu2009}.

Since the introduction of Airy beams (AiBs) by Siviloglou and Christodoulides \cite{Siviloglou2007, Sivilo2007}, in recent years it has attracted great interest in optics and atomic physics due to their unusual features such as the ability to remain diffraction-free over long distances while they freely accelerate during propagation. Most recently, several authors have addressed the propagation properties of AiBs superimposed with an OV due to its novel properties and potential applications \cite{Mazilu2009,Dai2010,Chen2015,Wei2018,Fang2018,Wu2018}.

On the other hand, several techniques can be used to generate experimentally the vortex beams \cite{Mazilu2009,Zhou2015,Fang2018,Wei2018}. The holographic techniques are now well-established techniques for generation of special optical beams (non-diffracting or diffraction-resistant) including OVs \cite{Vieira2012,Vieira2014,Vieira2015,Suarez2016,Yang2016,Fang2018}.  In fact, the computational holography technique with the use of computer-generated holograms and with the development of spatial light modulators (SLMs) has been able to reproduce efficiently many types of optical beams which are difficult to obtain using conventional methods \cite{Vieira2012,Vieira2014,Vieira2015,Suarez2016}. Also, the digital holography technique has been shown to be effective for analysis of optical wave and optical beam wavefronts \cite{Yepes2019}.  

In this work, we propose a novel and an accessible method to generate an array of Airy-vortex beams (AiVBs) from a superposition of N identical AiBs symmetrically located with respect to the propagation axis, where each beam is superimposed with OV and are characterized by the same topological charge $l$. We present a single experimental setup to obtain the beam intensity through the implementation of amplitude computer-generated holograms (CGHs) in spatial light modulators (SLM) and the phase analysis by digital holography (DH). These results are in agreement with the corresponding theoretical analytical solutions and they present excellent prospects for applications in optical manipulation techniques. \\

\h {\em\bf 2. Theoretical background} --- At the plane ($z=0$), the field distribution of $2D$ Airy beams propagating with finite energy and superimposed by an $l$-order OV can be expressed as \cite{Sivilo2007,Dai2010},
\begin{equation}
\begin{split}
\psi(s_{x},s_{y},\xi=0)&=\operatorname{Ai}(s_{x})\operatorname{Ai}(s_{y})\exp [a(s_{x} + s_{y})]\\
	& \times [( s_{x}-s_{x_{d}})  + i ( s_{y}-s_{y_{d}})]^l\,,  
\end{split}
\label{Airy}
\end{equation}
where $\left( s_{x}=x/x_{0}, s_{y}=y/y_{0} \right)$ is dimensionless transverse coordinates, $\xi=z/kw_{0}^{2}$ is a normalized propagation distance, $\operatorname{Ai}$ is the Airy function, $a$ is a positive quantity to guaranty finite energy, and $w_{0}$ is an arbitrary transverse scale. $s_{x_{d}}=x_{d}/w_{0}$ and $s_{y_{d}}=y_{d}/w_{0}$ where $x_{d}$ and $y_{d}$ denote the displacement of the singularity from the origin along the $x$-$y$ axes. 

The scalar field $\psi\left(x,y,z\right)$ can be calculated through Huygens-Fresnel integral which  determines the field at a distance $z$ as a function of the field at $z=0$, that is \cite{Morris2007,Besieris2016}.
\begin{equation}
\begin{split}
\psi(x,y,z)&=\dfrac{k}{2\pi z}\int \int \exp\left\lbrace \dfrac{i k}{2z}\left[\left(x_{1}-x \right)^2 + \left(y_{1}-y \right)^2 \right]  \right\rbrace   \\
	& \times  \psi\left(x_{1},y_{1}\right)  dx_{1} dy_{1}\,.
\end{split}
\label{HF}
\end{equation}

We consider $N$ AiVBs, through a $\theta=2\pi/N$ degree rotation of Eq.~\ref{Airy} in the transverse plane. We obtain a set of $N$ rotated AiVBs which accelerate mutually in opposite directions \cite{Lu2017}. The field of the resulting AiVBs can be described as 
\begin{equation}
\Psi(s_{x},s_{y},\xi)=\sum_{j=1}^{N} \psi_{j}(s_{x_{j}},s_{y_{j}},\xi)\,,
\label{AAB}
\end{equation}
where $\psi_{j}$ is the $j$th AiVB, $N$ is the number of  beams in the superposition and $(s_{x_{j}},s_{y_{j}})$ are the transversal coordinates of the $j$th AiVB and are given by
\begin{equation}
s_{x_{j}}=s_{x}\text{cos}\theta_{j}-s_{y}\text{sin}\theta_{j}, \quad s_{y_{j}}=s_{x}\text{sin}\theta_{j}+s_{y}\text{cos}\theta_{j}
\label{Rotation}
\end{equation}
the angle $\theta_{j} = 2(j-1)/N$ denotes the angle of rotation around the z-axis. For $l=1$, the analytical complex field distribution of the $j$th AiVBs can be expressed as
\begin{equation}
\psi_{j}(s_{x_{j}},s_{y_{j}},\xi)=\text{exp}[Q_{j}(s_{x_{j}},s_{y_{j}},\xi )][P_{j1}+P_{j2}+P_{j3}] \,,
\label{AAB}
\end{equation}
where
\begin{equation} 
Q(s_{x_{j}},s_{y_{j}},\xi) = (B_{jx}+B_{jy})+i(C_{jx}+C_{jy})\,, 
\label{Q}
\end{equation}
\begin{equation}
\begin{split} 
P_{j1}=Ai(D_{jx})Ai(D_{jy}) & (F_{jx} + F_{jy})\,,
\end{split}
\label{P1}
\end{equation}
\begin{equation} 
P_{j2}= - \xi  Ai(D_{jx})[Ai(1,D_{jy})+a Ai(D_{jy})]\,,
\label{P2}
\end{equation}
\begin{equation} 
P_{j3}= i\xi Ai( D_{jy})[Ai(1,D_{jx})+a Ai(D_{jx})]\,,
\label{P2}
\end{equation}
where $Ai(1,D_{jm})$ is the first derivative of the Airy function. $B_{jm}=a(s_{jm}-\xi/2 )$, $D_{jm}=s_{jm}-\xi^2/4 + ia\xi$, $C_{jm}=[-\xi^3/12+(a^2+s_{jm})\xi/2]$, and $F_{jm}=s_{jm}-\dfrac{\xi^2}{2} - s_{jm_{d}} $ for $m=x,y$. The component $P_{j1}$, represents a $j$th conventional AiVB after propagating a distance $z$. The term $P_{j2}$ denotes the profile of the $j$th AiB with the $y$ component described by the combination of Airy function and its derivative. $P_{j3}$ is similar to $P_{j2}$ with an exchange of the $x$ and $y$ coordinates, and a phase difference of $\pi/2$ \cite{Dai2010,Chen2015}.

\begin{figure}[H]
 \centering
 \includegraphics[scale=0.64]{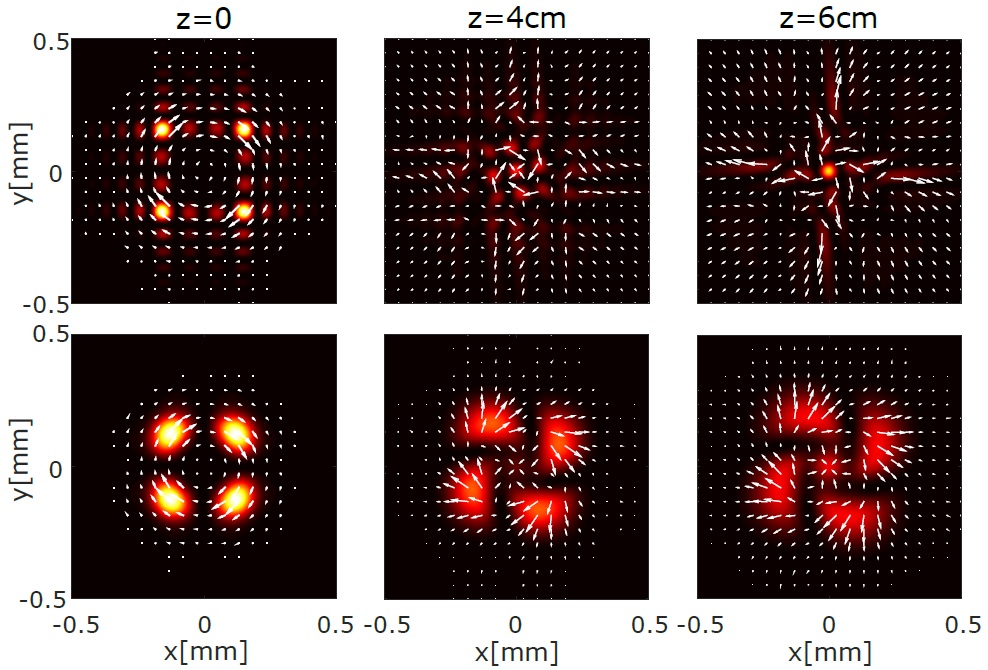}
\caption{Transverse Poynting vector $s_{\perp}=s_{x}+s_{y}$ (white arrows), for a $l=2$ superposition of $N=4$ AiVBs at $z=0$,$4$,$6$ cm. For $a=0.5$ (top row) and $a=1$ (bottom row), respectively, with the cross-section intensity as the background.}
\label{Poynting}
\end{figure}

In Figure~\ref{Poynting} illustrate the characteristics of this type of optical vortex beam originating from  an array of Airy-vortex beams (AiVBs) from a superposition of N identical AiBs symmetrically located with respect to the propagation axis, where each beam is superimposed with OV and are characterized by the same topological charge l. The Figure~\ref{Poynting} shows the cross-section intensity distribution (background intensity) of the AiVBs with $N=4$ and topological charge $l=2$ for $a=0.5$ (top row) and $a=1$ (bottom row) for different planes along of the propagation direction. This is obtained by transverse Poynting vector $s_{\perp}=s_{x}+s_{y}$ \cite{Sztul2008}. The direction and magnitude of the arrows (white) represent the direction and magnitude of the energy flow in the transverse plane. Initially, the energy flow exhibits clockwise rotation and change dynamically during propagation while each AiVBs is accelerated in the opposite direction with a slight deviation from the conventional AiBs \cite{Dai2010}. Note, that the energy density in the central region increases with the propagation distance, which is due to the process of auto-focusing of the fields. On the other hand, the density energy in the outer region for $a=1$ is less than that for $a=0.5$, due to different finite energy parameters of the AiVBs. \\

\h {\em\bf 3. Experimental generation and analysis of Airy-vortex beams via holographic methods} 

\h {\em 3.1. Optical reconstruction of Airy-vortex beams using CGH via SLM} --- The holographic computational methods are now a well established technique for generation and characterization of special optical beams and structured light, particularly non-diffracting beams and with orbital angular momentum. The computer-generated holograms of these special beams are calculated and implemented in spatial light modulators and reconstructed optically in a holographic setup. Particularly, these methods have generated experimental results of high quality and fidelity optical beams reconstruction compared to the theoretical predicted, because the holographic technique is an extremely accurate tool in the reconstruction of amplitude and phase of optical waves. 

Particularly, in this work to generate experimentally via a computational holographic technique the AiVBs we built a CGH from the Eq.~\ref{AAB}, which is optically reconstructed using an SLM  \cite{Vieira2015,Suarez2016,Arrizon2005}. The computer-generated hologram is calculated using an amplitude function which consists in varying the coefficient of transmission or reflection of the medium  from the following amplitude function
\begin{equation}
H\left(x,y\right)= \frac{1}{2}\left\lbrace \beta\left( x,y\right) +a\left( x,y\right)cos\left[\phi\left( x,y\right)-2\pi \left(\xi x + \eta y \right)  \right] \right\rbrace \,,
\label{transmission}
\end{equation}
where $a\left( x,y\right)$ is the amplitude and $\phi\left( x,y\right)$ is phase of the complex field, $\left(\xi,\eta \right)$ is a spacial frequency of the plane wave using as reference and $\beta\left( x,y\right)=\left[1+a^{2}\left(x,y \right)\right]/2 $ is the function bias taken as a soft envelope of the amplitude $a\left( x,y\right)$. The plane wave of reference is off-axis and introduces frequencies that separate the different orders of the encoded field.

\h {\em 3.2. Phase analysis of Airy-vortex beams using Digital Holography} --- The characterization of the phase distribution of this type of special beam has been made interferometric and holographic techniques \cite{Yepes2019,Khajavi2018,Brito2013,Gesualdi2008,Gesualdi2010,yepes2017dynamic}. Particularly in this work, the phase distribution was made via digital holography using the angular spectrum method. 

The computational reconstruction of the digital holograms of phase and intensity of the non-diffracting beams can be made by the Angular Spectrum method, where $A(k_{\xi},k_{\eta},z=0)$ is the  angular spectrum of the hologram is obtained \cite{Yepes2019}. The angular spectrum is defined as the Fourier Transform of the digital hologram:

\begin{equation}
A(k_{\xi},k_{\eta},z=0)={\int \int {E(\xi_{0},\eta_{0},0)} {\exp{[-i(k_{\xi}{\xi_{0}}+k_{\eta}{\eta_{0}})]}} {d\xi_{0} d\eta_{0}}}\,
\label{SA}
\end{equation}

where, $k_{\xi}$ and $k_{\eta}$ are the spatial frequencies of ${\xi}$ and ${\eta}$. $E(\xi_{0},{\eta_{0}},0)$ is the complex amplitude of the digital hologram. 

Through an inverse Fourier Transform, the field correspondent to this plane is calculated $E_{H}(x,y)$. This calculation results in a matrix of complex numbers and the phase of the object wave can be determined \cite{Yepes2019,Khajavi2018,Brito2013}: 

\begin{equation}
\Phi(x,y)={\arctan \frac{\emph{I}[E_{H}(x,y)]}{\emph{R}[E_{H}(x,y)]}},
\label{PH}
\end{equation}

\begin{figure}[H]
 \centering
 \includegraphics[scale=0.60]{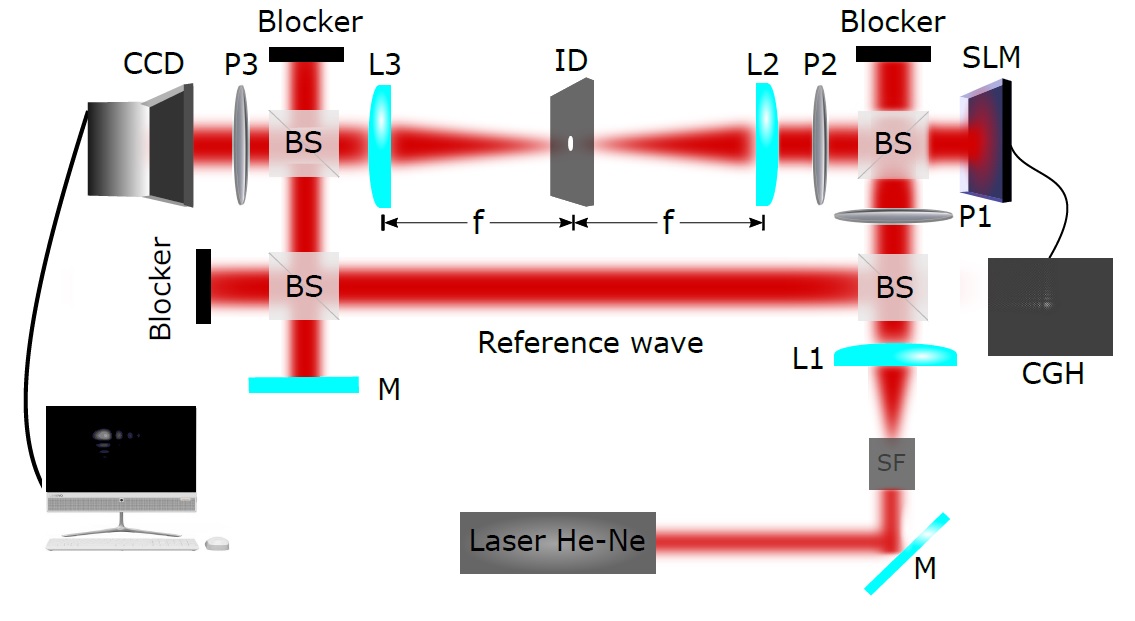}
\caption{Experimental holographic setup for optical generation via CGHs and analysis of phase distribution using DH. Where the laser is He-Ne $(632.8nm)$; $SF$ is a spatial filter; $L1$, $L2$, and $L3$ are lens; $\text{P}_{1}$, $\text{P}_{2}$ and $\text{P}_{3}$ are polarizers; $M$ are mirrors; $BS$ are beam-splitter; SLM is a reflective spatial light modulator; ID is the mask; and CCD is the camera for image acquisition.}
\label{Setup}
\end{figure}

\h {\em 3.3. Experimental setup} --- The Fig.~\ref{Setup} shows the experimental holographic setup for generation and characterization of AiVBs. 

In the generation of the fields, the recording and reconstruction process are made numerically and optically, respectively. The He-Ne laser beam $\left( \lambda=632.8nm\right)$, is expanded by the spatial filters $SF$ and collimated by the lens $L1$, incident perpendicularly on the surface of the SLM (Model $LC-R1080$, Holoeye Photonics), with a display of $1980\times 1200$ and pixel size $8.1\mu m$) which is placed at the input plane (focus of lens $L2$). To obtain an amplitude modulation, the optical axis of polarizer $P1$ is aligned forming an angle of $0^{\circ}$ and of the polarizer $P2$ in $90^{\circ}$ with relation to the axis $y$ of the SLM. A $4f$ system composed of two lenses ($L2$ and $L3$) allows select the correct diffraction order of the reconstructed (diffracted) beam. The mask $ID$, placed at the Fourier plane, is used to block the undesired higher diffraction orders \cite{Vieira2015,Suarez2016}. Finally, the desired beam is acquired by a CCD camera which is parallel-aligned to the propagation axis of the beam. 

And, for phase analysis by DH, the recording and reconstruction process is made optically and numerically respectively. In this case, for an optical recording of the holograms we coupled the experimental apparatus for the generation of optical beams to a Mach-Zehnder interferometer using a beam-splitter $(BS)$ at the output of the lens $L_{1}$ (the mirror $M$ guarantees the same optical path difference in the interferometer), which results in the digital hologram. The optical beams are finally reconstructed numerically using the digital hologram obtaining information of its phase distribution. The computational reconstruction of the digital holograms of phase and intensity of the non-diffracting beams can be made by the Angular Spectrum method, where  the angular spectrum of the hologram is obtained, see \cite{Yepes2019}. \\

\h {\em\bf 4. Results and Discussion} --- For optical generation of Airy beams was used the experimental setup Fig.~\ref{Setup}. Initially, we build a CGH of the field described by Eq.~\ref{AAB} from the Eq.~\ref{transmission} adopting the carrier of frequencies $\eta=\xi=\Delta p/5$ for the plane wave of reference, where $\Delta p=1/\delta p$ is the bandwidth and $\delta p$ is the individual pixel size ($8.1\mu m$). \\

As an example for this method, consider a superposition of AiVBs formed by $N$ AiVBs where each vortex propagates along the $z$-axis with the same topological charge $l$ and characterized by the following parameters: $\lambda=632.8nm$, $a=1$, $w_{0}=50\mu m$ and $x_{d}=y_{d}=0$. We illustrate three examples below. \\

\begin{figure}[H]
 \centering
 \includegraphics[scale=0.84]{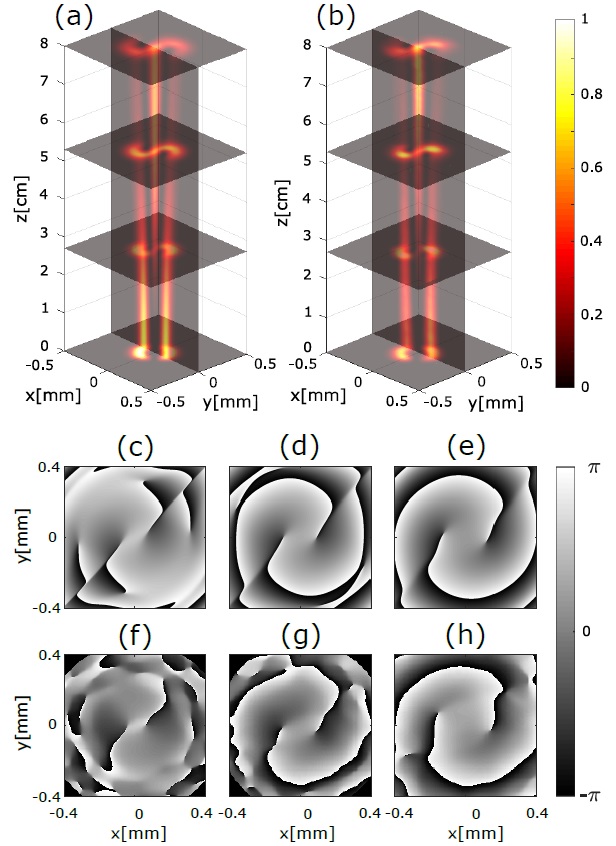}
\caption{Dynamic propagation of the intensity distribution for a superposition of AiVBs $(N=2)$ with $l=1$ in $(a)$ theoretical and $(b)$ experimental results. The transverse cross-section represent the planes $z=0$, $z=2.6cm$, $z=5.4cm$, and $z=8.0cm$. $(c)$-$(e)$ are the theoretical phase distribution and $(f)$-$(h)$ are the experimental results.}
\label{VAAB_l1_N2}
\end{figure}

\textbf{First example ($N=2$; $l=1$):} A superposition with two AiVBs and each vortex have topological charge $l=1$. The dynamic propagation of the normalized intensity distribution can be simulated as shown in Fig.~\ref{VAAB_l1_N2}$(a)$. The corresponding experimental result, obtained by optical reconstruction of CGH, shown in Fig.~\ref{VAAB_l1_N2}$(b)$, agrees with the theoretical predictions. Two petals of the intensity profile, generated by the superposition of the AiVBs can be observed. In Figs.~\ref{VAAB_l1_N2}$(c)$-$(e)$ the theoretical phase distribution taken at the respective locations (except for $z=0$), can be compared to the experimental results obtained numerically by digital holography in Figs.~\ref{VAAB_l1_N2}$(f)$-$(h)$. 

\begin{figure}[H]
 \centering
 \includegraphics[scale=0.84]{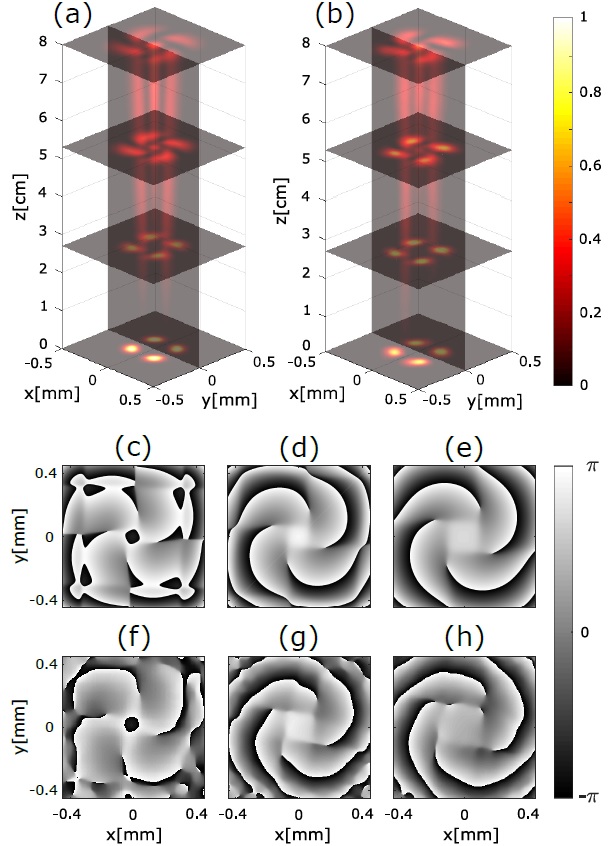}
\caption{Dynamic propagation of the intensity distribution for a superposition of AiVBs $(N=4)$ with $l=2$ in $(a)$ theoretical and $(b)$ experimental results. The transverse cross-section represent the planes $z=0$, $z=2.6cm$, $z=5.4cm$, and $z=8.0cm$. $(c)$-$(e)$ are the theoretical phase distribution and $(f)$-$(h)$ are the experimental results.}
\label{VAAB_l2_N4}
\end{figure}

\textbf{Second example ($N=4$; $l=2$):} A superposition with four AiVBs and each vortex have topological charge $l=2$. The dynamic propagation of the normalized intensity distribution theoretical and experimental can be observed in Figs.~\ref{VAAB_l2_N4} $(a)$ and $(b)$. The theoretical phase distribution taken at the respective locations (except for $z=0$) Figs.~\ref{VAAB_l2_N4} $(c)$-$(e)$, can be compared to the experimental results for Figs.~\ref{VAAB_l2_N4} $(f)$-$(h)$.

\begin{figure}[H]
 \centering
 \includegraphics[scale=0.84]{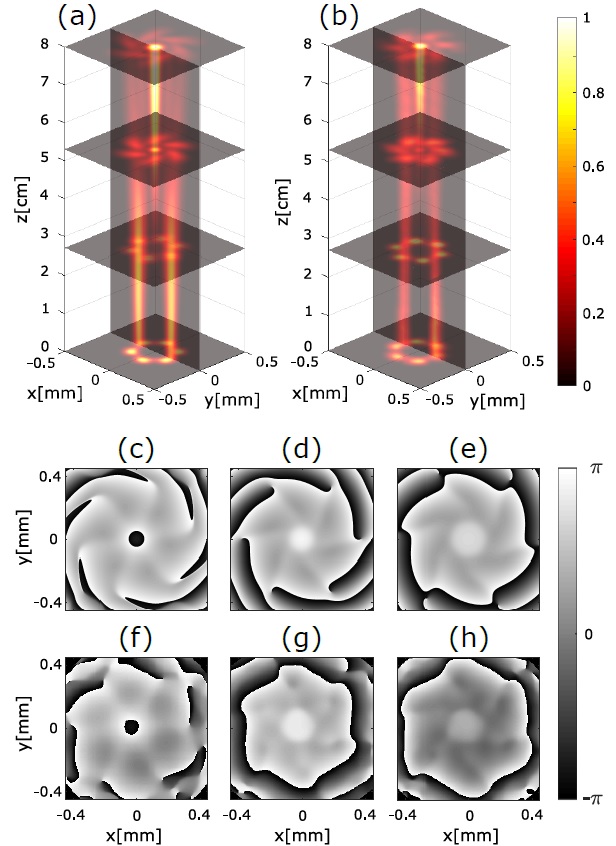}
\caption{Dynamic propagation of the intensity distribution for a superposition of AiVBs $(N=6)$ with $l=2$ in $(a)$ theoretical and $(b)$ experimental results. The transverse cross section represent the planes $z=0$, $z=2.6cm$, $z=5.4cm$, and $z=8.0cm$. $(c)$-$(e)$ are the theoretical phase distribution and $(f)$-$(h)$ are the experimental results.}
\label{VAAB_l2_N6}
\end{figure}

\textbf{Third example ($N=6$; $l=2$):} Lastly, a superposition with six AiVBs and each vortex have topological charge $l=2$. Fig.~\ref{VAAB_l2_N6} $(a)$-$(b)$ and \ref{VAAB_l2_N6} $(c)$-$(h)$ shows both the theoretical and experimental results of the dynamic propagation of the normalized intensity and phase distribution respectively.

For the three illustrated examples of AiVBs, the experimental results are in agreement with those predicted theoretically. Also, note that the gap between the intensity peaks, which is formed by OV superimposed, become larger with the increase of the topological charge, as well as, the number of the petals on the intensity profile with the number of the AiVBs. However, the center region of null intensity disappears with the propagation distance. Moreover, the phase distribution analysis allows to determine the position of each vortex (singularity point) and the sign of the topological charge is determined by the handedness in which the phase increases. Note that each vortex beams in the superposition rotates in a counterclockwise direction, where the speed and sense rotation depends on the topological charge \cite{Dai2010,Fang2018}, and its separation increases with the propagation distance due to the trajectory which occur in the opposite direction, and which is similar to the parabolic trajectory of the AiBs \cite{Dai2010}.

This work presents a novel, straightforward and accessible method to generate an array of Airy-vortex beams (AiVBs) from a superposition of AiBs whereby each beam is superimposed with OV of the same topological charge $l$. The experimental results are in agreement with the theoretical predictions and the  interesting goals of this work was the use of this technique for the complete experimental generation and characterization of intensity and phase of the array of Airy-vortex beams depicted in Figs.~\ref{VAAB_l1_N2},~\ref{VAAB_l2_N4} and ~\ref{VAAB_l2_N6}. The present work opens an exciting possibility for simultaneous generation and analysis of these beams for applications in optical guiding and trapping of particles. Whereby several optical traps could be generated simultaneously, which would be suitable for trapping and rotating micro-particles, where the phase defines the direction of the moving particles.  \\

\h {\em\bf 4. Conclusions} --- In summary, this work presents a novel, straightforward and accessible method to generate an array of Airy-vortex beams (AiVBs) from a superposition of AiBs whereby each beam is superimposed with OV of the same topological charge $l$. Additionally, a single experimental apparatus is proposed allowing both the optical generation of an array of AiVBs via optical reconstruction by CGH using an SLM; and, their phase analysis, by a DH technique. The experimental results are in agreement with the theoretical predictions and the  interesting goals of this work was the use of this technique for the complete experimental generation and characterization of intensity and phase of the array of Airy-vortex beams. The present work opens an exciting possibility for simultaneous generation and analysis of these beams for applications in optical guiding and trapping of particles. Whereby several optical traps could be generated simultaneously, which would be suitable for trapping and rotating micro-particles, where the phase defines the direction of the moving particles. Further, others potential interesting scientific and technological applications as in optical communications, through the phase shape is possible identify the number and order of orbital angular momentum used for multiplex information and increase data transmission; and, phase variations in optical metrology. \\

\h {\em Acknowledgments} The authors acknowledge financial support from UFABC, CAPES, FAPESP (grant 16/19131-6) and CNPq (grant 302070/2017-6).


\begin{thebibliography}{99}

\bibitem{Allen1992}
L.~Allen, M.~W. Beijersbergen, R.~Spreeuw, and J.~Woerdman, Orbital angular momentum of light and the transformation of laguerre gaussian laser modes, Physical Review A 45 8185 (1992) 8185--8190.


\bibitem{Paterson2001}
L.~Paterson, M.~MacDonald, J.~Arlt, W.~Sibbett, P.~Bryant, and K.~Dholakia,Controlled rotation of optically trapped microscopic particles, Science 292 (2001) 912--914.

\bibitem{Dholakia2011}
K.~Dholakia and T.~{\v{C}}i{\v{z}}m{\'a}r, Shaping the future of manipulation, Nature Photonics 5 (2011) 335--343.

\bibitem{Willner2015}
A.~E. Willner, H.~Huang, Y.~Yan, Y.~Ren, N.~Ahmed, G.~Xie, C.~Bao, L.~Li, Y.~Cao, Z.~Zhao, Optical communications using orbital angular momentum beams, Advances in Optics and  Photonics 7 (2015) 66-106.

\bibitem{Molina2007}
G.~Molina-Terriza, J.~P. Torres, and L.~Torner, Twisted photons, Nature Physics 3 (2007) 305--311.

\bibitem{Flossmann2005}
F.~Flossmann, U.~Schwarz, and M.~Maier, Propagation dynamics of optical vortices in laguerre gaussian beams, Optics Communications 250 (2005) 218--230.

\bibitem{Orlov2002}
S.~Orlov, K.~Regelskis, V.~Smilgevi{\v{c}}ius, and A.~Stabinis, Propagation of bessel beams carrying optical vortices, Optics Communications 209 (2002) 155--165.

\bibitem{Mazilu2009}
M.~Mazilu, J.~Baumgartl, T.~{\v{C}}i{\v{z}}m{\'a}r, and K.~Dholakia, Accelerating vortices in airy beams, Proc. of SPIE 7430 (2009) 74300C-74304C.

\bibitem{Siviloglou2007}
G.~A. Siviloglou, J.~Broky, A.~Dogariu, D.~N. Christodoulides, Observation of accelerating airy beams, Physical Review Letters 36 (2007) 1760--1765.

\bibitem{Sivilo2007}
G.~A. Siviloglou, D.~N. Christodoulides, Accelerating finite energy airy beams, Optics Letters 32 (2007) 979--981.

\bibitem{Dai2010}
H. T. Dai, Y. J. Liu, D. Luo, and X. W. Sun, Propagation dynamics of an optical vortex imposed on an airy beam, Optics Letters 35 (2010) 4075--4077.

\bibitem{Chen2015}
B.~Chen, C.~Chen, X.~Peng, and D.~Deng, Propagation of airy gaussian vortex beams through slabs of right-handed materials and left-handed materials, Optical Society of America 32 (2015) 173--178.

\bibitem{Wei2018}
B.-Y. Wei, S.~Liu, P.~Chen, S.-X. Qi, Y.~Zhang, W.~Hu, Y.-Q. Lu, and J.-L. Zhao, Vortex airy beams directly generated via liquid crystal q-airy-plates, Applied Physics Letters 112 (2018) 121101-121108.

\bibitem{Fang2018}
Z.-X. Fang, Y.~Chen, Y.-X. Ren, L.~Gong, R.-D. Lu, A.-Q. Zhang, H.-Z. Zhao, and P.~Wang, Interplay between topological phase and selfacceleration in a vortex symmetric airy beam, Optics Express 26 (2018) 7324--7335.

\bibitem{Wu2018}
Y.~Wu, L.~Shao, and J.~Nie, Evolution dynamics of vortex quasi-airy beams, JOSA B 35 (2018) 972--979.

\bibitem{Zhou2015}
J.~Zhou, Y.~Liu, Y.~Ke, H.~Luo, and S.~Wen, Generation of airy vortex and airy vector beams based on the modulation of dynamic and geometric phases, Optics Letters 40 (2015) 3193--3196.

\bibitem{Vieira2012}
T.~A. Vieira, M.~R.~R. Gesualdi, M.~Zamboni-Rached, Frozen waves: experimental generation, Optics Letters 37 (2012) 2034--2036.

\bibitem{Vieira2014}
T.~A. Vieira, M.~Zamboni-Rached, M.~R.~R. Gesualdi, Modeling the spatial shape of nondiffracting beams: Experimental generation of Frozen Waves via holographic method, Optics Communications 315 (2014) 374--380.

\bibitem{Vieira2015}
T.~A. Vieira, M.~R.~R. Gesualdi, M.~Zamboni-Rached, E.~Recami, Production of dynamic frozen waves: controlling shape, location (and speed) of diffraction resistant beams, Optics Letters 40 (2015) 5834--5837.

\bibitem{Suarez2016}
R.~A.~B. Suarez, T.~A. Vieira, I.~S.~V. Yepes, M.~R.~R. Gesualdi,   Photorefractive and computational holography in the experimental generation of Airy beams, Optics Communications 366 (2016) 291--300.

\bibitem{Yang2016}
Y.~Lu, B.~Jiang, S.~Lu, Y.~Liu, S.~Li, Z.~Cao, and X.~Qi, Arrays of gaussian vortex bessel and airy beams by computer-generated hologram, Optics Communications 363 (2016) 85--90.

\bibitem{Gesualdi2008}
M.~R.~R. Gesualdi, M. Muramatsu, E.~A. Barbosa, Light-induced lens analysis in photorefractive crystals employing
 phase-shifting real time holographic interferometry, Optics Communications 281 (2008) 5739--5744.

\bibitem{Gesualdi2010}
M.~R.~R. Gesualdi, D. Soga, M. Muramatsu, R.~D. Paiva Junior, Wave optics analysis by phase-shifting real-time holographic interferometry, Optik 121 (2010) 80--88.

\bibitem{yepes2017dynamic}
I.~S. Yepes, M.~R. Gesualdi, Dynamic digital holography for recording and reconstruction of 3d images using optoelectronic devices, Journal of Microwaves, Optoelectronics and Electromagnetic Applications 16~(3) (2017) 801--815.

\bibitem{Brito2013}
I.~V. Brito, M.~R.~R. Gesualdi, J. Ricardo, F. Palacios, M. Muramatsu, J.~L. Valin, Photorefractive digital holographic microscopy applied in microstructures analysis, Optics Communications 286 (2013) 103--110.

\bibitem{Yepes2019}
I.~S. Yepes, T.~A. Vieira, R.~A. Suarez, S.~R. Fernandez, M.~R. Gesualdi, Phase and intensity analysis of non-diffracting beams via digital holography, Optics Communications 437 (2019) 121--127.

\bibitem{Morris2007}
J.~E. Morris, M.~Mazilu, J.~Baumagartl, T.~{\v{C}}i{\v{z}}m{\'a}r, K.~Dholakia, Propagation characteristics of airy beams: Dependence upon spatial coherence and wavelength, Optics Express 17 (2009) 13236--13245.

\bibitem{Besieris2016}
I.~M. Besieris, A.~M. Shaarawi, and M.~Zamboni-Rached, Accelerating airy beams in the presence of inhomogeneities, Optics Communications 369 (2016) 279--288.

\bibitem{Lu2017}
Q.~Lu, S.~Gao, L.~Sheng, J.~Wu, Y.~Qiao, Generation of coherent and incoherent airy beam arrays and experimental comparisons of their scintillation characteristics in atmospheric turbulen, Applied Optics 56~(13) (2017) 3750--3757.

\bibitem{Sztul2008}
H.~Sztul, R.~Alfano, The poynting vector and angular momentum of airy beams, Optics Express 16 (2008) 9411--9416.


\bibitem{Arrizon2005}
V.~Arrizon, G.~Mendez, D.~Sanchez de~la llave, Accurate encoding of arbitrary complex fields with amplitude only liquid crystal spatial ligth modulators, Optics Express 13 (2005) 7913--7929.

\bibitem{Khajavi2018}
B.~Khajavi, J.~Ureta, and E.~Galvez, Determining vortex-beam superpositions by shear interferometry, Photonics 5 (2018) 16-24.



\end{thebibliography}
\end{document}